\documentclass[paper]{aastex61}

\usepackage{url}
\usepackage{tabularx}
\usepackage{amsmath}
\usepackage{rotating}
\usepackage{graphicx}

\shorttitle{Cyclic Changes of the Sun's Seismic Radius}
\shortauthors{Kosovichev \& Rozelot}

\begin{document}
	
	\title{Cyclic Changes of the Sun's Seismic Radius}
	
	\correspondingauthor{Alexander Kosovichev}
	\email{alexander.g.kosovichev@njit.edu}

	\author{Alexander Kosovichev}
	\affiliation{Center for Computational Heliophysics, New Jersey Institute of Technology, Newark, NJ 07102, USA}
	\affiliation{Department of Physics, New Jersey Institute of Technology, Newark, NJ 07102, USA}

\author{Jean-Pierre Rozelot}
\affiliation{Nice University, OCA-CNRS, 77 Chemin des Basses Mouli\`eres, 06130 Grasse, France}

\begin{abstract}

The questions whether the Sun shrinks with the solar activity and what causes this have been a subject of debate. Helioseismology provides means to measure with high precision the radial displacement of subsurface layers, co-called `seismic radius', through analysis of oscillation frequencies of surface gravity (f) modes. Here, we present results of a new analysis of twenty one years of helioseismology data from two space missions, Solar and Heliospheric Observatory (SoHO) and Solar Dynamics Observatory (SDO), which allow us to resolve previous uncertainties and compare variations of the seismic radius in two solar cycles. After removing the f-mode frequency changes associated with the surface activity we find that the mean seismic radius is reduced by 1-2~km during the solar maxima, and that most significant variations of the solar radius occur beneath the visible surface of the Sun at the depth of about $5\pm 2$~Mm, where the radius is reduced by 5-8~km. These variations can be interpreted as changes in the solar subsurface structure caused by  predominately vertical $\sim 10$~kG magnetic field.     
\end{abstract}

\section{Introduction}
Accurate measurement of the solar radius and its variations is one of the oldest astronomical problems. It is important for two primary reasons. First, the radius serves as an astronomical standard. Second, the solar radius and, more generally, shape variations reflect still poorly understood physical processes associated with the cyclic magnetic activity, which occur in the Sun's interior and affect the surface. First accurate measurements of the size of the Sun performed in 18th and 19th centuries indicated that ``the systematically larger diameters correspond to the time when the number of spots and protuberances is lower"  \citep{Secchi1872,Auwers1873}. \citet{Secchi1872} conjectured that ``the  effect  of  the active   forces  in  the  Sun,   which   are   made   known  to us by the variable formations on its surface, may produce  changes of volume in the masses of luminous gas, perhaps perceptible in  accurate observations of the  Sun's diameter." Modern theories motivated by measurements of irradiance variations and shifts in solar oscillation frequencies during the solar cycles attempt to explain these measurements in terms changes in the structure of the Sun caused by large-scale and turbulent magnetic fields  \citep{Goldreich1991,Dziembowski2004,Mullan2007}.  

The long standing question whether the solar radius is constant or not is still debated. There have been numerous studies using spacecraft and ground-based instruments that led in the past to conflicting results \citep[for a review, see ][]{Rozelot2016}.
The solar radius is determined by the position of the inflection point of the limb brightness. Thus, measuring with a high accuracy the diameter of the Sun is a challenge at the cutting edge of modern techniques \citep{Bush2010}.  Most recent measurements of the solar limb from the PICARD satellite put an upper limit of 14.5~km (twenty parts in a million) on the solar radius changes during the rising phase of the current sunspot cycle \citep{Meftah2015}. 

Helioseismology provides an alternative measure of the solar radius, so-called ``seismic radius'' \citep{Schou1997}. The seismic radius is determined from frequencies of surface gravity waves (f-modes). The f-mode frequencies depend on the local gravity acceleration and the oscillation wavelength which in turn depends on the solar radius and the mode spherical harmonic angular degree. The oscillation frequencies are measured to a very high precision ($\sim 10^{-6}$), and provide an accurate measure of the seismic radius. The surface gravity waves travel beneath the visible surface of the Sun, and their frequencies are sensitive to the sharp density gradient in the near-surface layers. Therefore, the seismic radius is an attribute of the subsurface stratification of the Sun. It is different from the visual radius; the relationship between them can be made only through modeling \citep{Sofia2005}. First helioseismology measurements \citep{Schou1997} showed that the Sun's seismic radius is about 300~km smaller than the predictions of the standard evolutionary solar model calibrated to the visual radius. The discrepancy was explained by an apparent shift of the limb inflection point due to light absorption in the solar atmosphere \citep{Haberreiter2008}. This led to a revision of the standard solar radius to the value of 695,700~km by the International Astronomical Union in 2015. 

Initial measurements of the seismic radius from SoHO showed that its changes do not exceed 1-2 km per year \citep{Dziembowski2001,Antia2003}. It was found that the helioseismic measurements have to take into account sensitivity of the f-mode frequencies to surface perturbations \citep{Dziembowski2001}, and also dependence of the radial displacement on depth \citep{Lefebvre2005}. In addition, the measurements are affected by periodic annual variations due to orbital motion of the Earth and spacecraft orbit around the Sun, and also potential changes in the instrumental sensitivity \citep{Antia2003}. Our new analysis which includes observational data from two missions and covers almost two solar cycles accounts for all these factors.

\section{Observational Data}
Here, we use helioseismology data obtained in 1996-2017 from two NASA missions: Solar and Heliospheric Observatory (SoHO) (1996-2010) and Solar Dynamics Observatory (SDO) (2010-2017). The data from the helioseismology  instruments, Michelson Doppler Imager (MDI) \citep{Scherrer1995} and Helioseismic and Magnetic Imager (HMI) \citep{Scherrer2012}, are available on-line from the SDO JSOC (Joint Science Operations Center) archive: http://jsoc.stanford.edu. The mode frequency analysis is performed using 72-day series of full-disk Dopplergrams. The HMI high-resolution data are specially prepared to match the spatial resolution of the MDI Medium-$\ell$ Structure Program  \citep{Kosovichev1997}, and perform uniform data processing from the two instruments \citep{Larson2016}. The frequency fitting of the oscillation power spectra was performed by using a symmetrical Gaussian line profile.  Compared to the frequencies fitted with an asymmetrical profile these frequencies have a systematic shift of about $0.006~\mu$Hz, which remains constant during the solar cycle. We use the frequency data determined from the symmetrical fits because these data are less noisy and more complete. Additional systematic errors can be due various instrumental effects, such as image-scale errors, cubic distortion from the instrument optics, misalignment of the CCD, an error in the inclination of the Sun’s rotation axis, and a potential tilt of the CCD. These errors and their corrections in the mode fitting procedure were described by \citet{Larson2018}. Even after the corrections these factors cause systematic 6-month variations of the f-mode frequencies. Most likely, these variations are caused by an error in the inclination of the Sun’s rotation axis. In our analysis, they cause 6-month or annual variations of the seismic radius. To eliminate these systematic errors we applied a smoothing Gaussian filter with the standard deviation of one year.

We study the whole time span ranging from April 30, 1996, to June 4, 2017. The MDI data cover the initial period until 03/20/2011, and the HMI data cover the rest. The total number of the frequency datasets combined from the two instruments for our analysis was 105. For all these periods we selected a common subset of f-modes. It includes 152 modes in the range of angular degree $\ell$ from 139 to 299. Thus, the mode sets used in our analysis are identical in all 105 data sets.  In addition to the mean f-mode frequencies that are sensitive to variations of the radial structure, the MDI and HMI observations provide data on frequency splitting, which allow us to investigate variations of the solar differential rotation and asphericity with the solar cycle \citep{Kosovichev2018}.

Figure 1 shows the f-mode frequency differences observed during the maximum of Solar Cycle 23 (observing interval with mid date 2 November 2001), the solar minimum between Cycles 23 and 24 (10 June 2009), and during the Cycle 24 maximum (15 May 2014), relative to the first observing 72-day interval with the mid date on 6 June 1996. The relative frequency difference reached about $7\times 10^{-5}$ in Cycle 23 and $5\times 10^{-5}$ in Cycle 24. During the activity minimum of 2009-10 the frequencies were lower than in the previous minimum of 1996-97, probably reflecting the lower solar magnetic activity during the last minimum. 

\begin{figure}
	\centering
	\includegraphics[width=10cm]{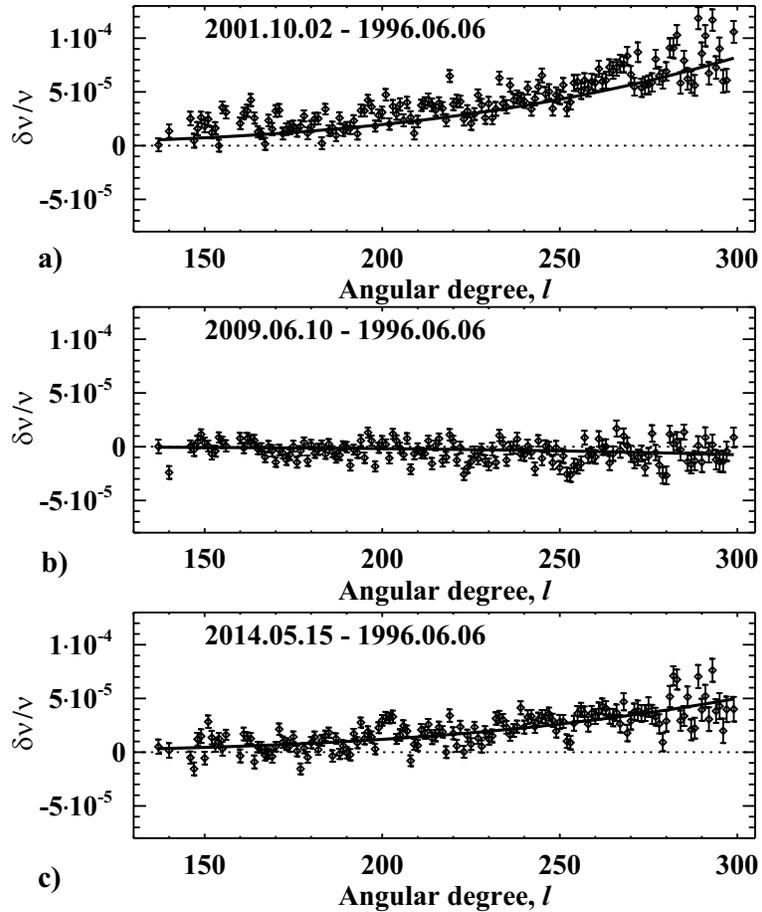}
	\caption{Frequencies of f-modes observed during various phases of the solar cycle: (a) the maximum of Cycle 23,  (b) the minimum between Cycles 23 and 24, and (c) the maximum of Cycle 24, relative to the frequencies observed during the first 72-day observing interval (the mid date of 6 June 1996 corresponds to the solar minimum between Cycles 22 and 23). The error bars are $\pm 1\sigma$. The solid curves show the surface term fits.}\label{fig1}
\end{figure}

\section{Theoretical Model}
For the data analysis we employ a theoretical model developed by Dziembowski et al. \citep{Dziembowski2001,Dziembowski2004, Dziembowski2005}. The model considers two primary contributions  to the f-mode frequencies, arising from changes in the stratification of subsurface layers and also from surface effects caused by  interaction of f-modes with surface magnetic fields and changes in the near-surface structure. Following this model we represent the relative frequency variations, ${\delta\nu_\ell}/{\nu_\ell}$, in the following form:
\begin{equation}
\frac{\delta\nu_\ell}{\nu_\ell}=-\frac{3}{2}\left(\frac{\Delta R}{R}\right)_\ell+\frac{\gamma_s}{I_\ell},
\label{eq1}
\end{equation}
where $(\Delta R/R)_\ell$ are relative variations of the seismic radius determined by f-modes of angular degree $\ell$,  and $\gamma_s$ is a parameter that describes variations of the surface effects with the solar cycle (so-called, the `surface term'), and $I_\ell$ is the mode inertia that takes into the dependence of this term on the mode angular degree.  The f-mode frequencies can be affected by a number other than the variations of the seismic radius, e.g. solar-cycle changes in turbulent convection and surface magnetic fields. These effects are scaled with the inverse mode inertia, and taken into account by the second term of Eq.~\ref{eq1} \citep{Dziembowski2001}.

The first right-hand-side term is calculated using the variational principle for nonradial stellar oscillations \citep{Chandrasekhar1964}, and provides a relationship between displacements of the subsurface layers, $\delta r/r$, and the frequency variations
in the following form:
\begin{equation} 
\left(\frac{\Delta R}{R}\right)_\ell=\frac{1}{I_\ell}\int_0^{R_\odot} S_\ell(r)\frac{\delta r}{r}dr,
\label{eq2}
\end{equation}
where the kernel function, $S_\ell(r)=\ell\rho(r) g(r)r^3|\vec{\xi}_\ell|^2/(2\pi\nu_\ell)^2$, describes the sensitivity of the f-mode frequencies to the radial displacement \citep{Dziembowski2004}, $I_\ell=\int_0^{R_\odot} |\vec{\xi}_\ell|^2\rho r^2dr$ in the mode mass, $\rho(r)$ in the mass density,  $g(r)$ is the gravity acceleration, $R_\odot$ is the Sun's photospheric radius, and $\vec{\xi}_\ell$ is the mode displacement eigenfunction.  Accuracy of the sensitivity kernels was tested using pairs of different solar models by \citet{Chatterjee2008}, who found that for high $\ell$ values additional terms that depend on variations of the sound speed near the surface may become significant, but for the medium-degree modes ($\ell < 300$) the displacement sensitivity kernel is dominant.  The primary difference in their test was due to large differences between the models close to the surface. In our model, the surface effects are taken into account by the second term in the RHS of Eq.~\ref{eq1}.

{  The surface term, $\gamma_s/I_\ell$,  is determined empirically by fitting it to the observed variations ${\delta\nu_\ell}/{\nu_\ell}$. Then, the first term of Eq.~\ref{eq1}, $-\frac{3}{2}\Delta R/R$, is determined by subtracting the surface term fit from the observed frequency variations,$\delta\nu_\ell/\nu_\ell$. To ensure that there is no residual correlation the surface term is represented in the form: $\gamma_s=-\frac{3}{2}\gamma_0S_\ell(R_\odot)$, where $S_\ell(R)$ is the surface values of the f-mode sensitivity function (see Eq.~\ref{eq2}), and $\gamma_0$ is the fitting coefficient.} Examples of this fitting are shown in Fig.~\ref{fig1} by solid curves. Apparently, the surface term fits the $\ell$-dependence quite well meaning that the main part of the frequency variations can be assigned to the 'surface effects' which can be due to both, variations of the solar structure and interaction of f-modes with magnetic fields at the solar surface. The surface term is subtracted from the observed frequency variations, and the remaining difference represents changes in the seismic radius described by Eq.~\ref{eq2} \citep{Dziembowski2001}. 
 \begin{figure}
 	\centering
 	\includegraphics[width=10cm]{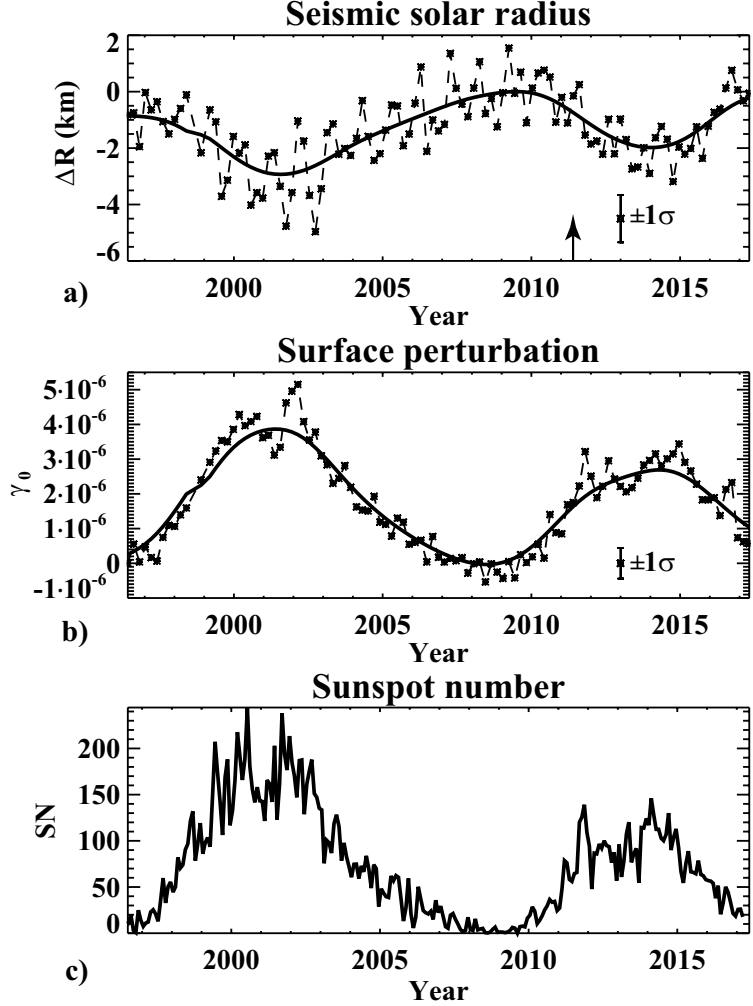}
 	\caption{Solar cycle variations of: a) the mean seismic radius, b) the coefficient, $\gamma_0$, of the surface perturbation of f-mode frequencies, and c) the sunspot number averaged for the same time intervals as the helioseismology data. The arrow in panel (a) indicates the start of the HMI data set.}\label{fig2}
 \end{figure}
 
 Figure~\ref{fig2}a shows the temporal variations of the seismic radius averaged over all $\ell$ values, relative to the seismic radius in 2009, during the solar minimum between Solar Cycles 23 and 24. We adopt this value as a reference. Figure~\ref{fig2}b shows the variations of the surface term coefficient, $\gamma_0$. It changes in phase with the solar activity and correlates quite well with the sunspot number (Fig.~\ref{fig2}c), except that it decays slower than the sunspot number in the declining phase of the solar cycles. The averaged seismic radius changes in anti-correlation, meaning that it becomes smaller during the solar maxima, in agreement with the earlier results  \citep{Dziembowski2001,Dziembowski2005,Lefebvre2007}. It shows annual variations, particularly, during the SoHO/MDI observations in 1996-2010. In the past, these variations prevented from making a definite conclusion whether the Sun `shrinks' with the solar activity  \citep{Antia2003}. The annual variations due to observing conditions are significantly reduced after 2010 when the SoHO/MDI observations were replaced by SDO/HMI.  The data show that the mean seismic radius decreased by about 3 km during the Cycle 23 maximum in 2000-2003, and by about 2 km during the Cycle 24 maximum in 2013-2015.

\section{Inversion results}

\begin{figure}
	\centering
	\includegraphics[width=10cm]{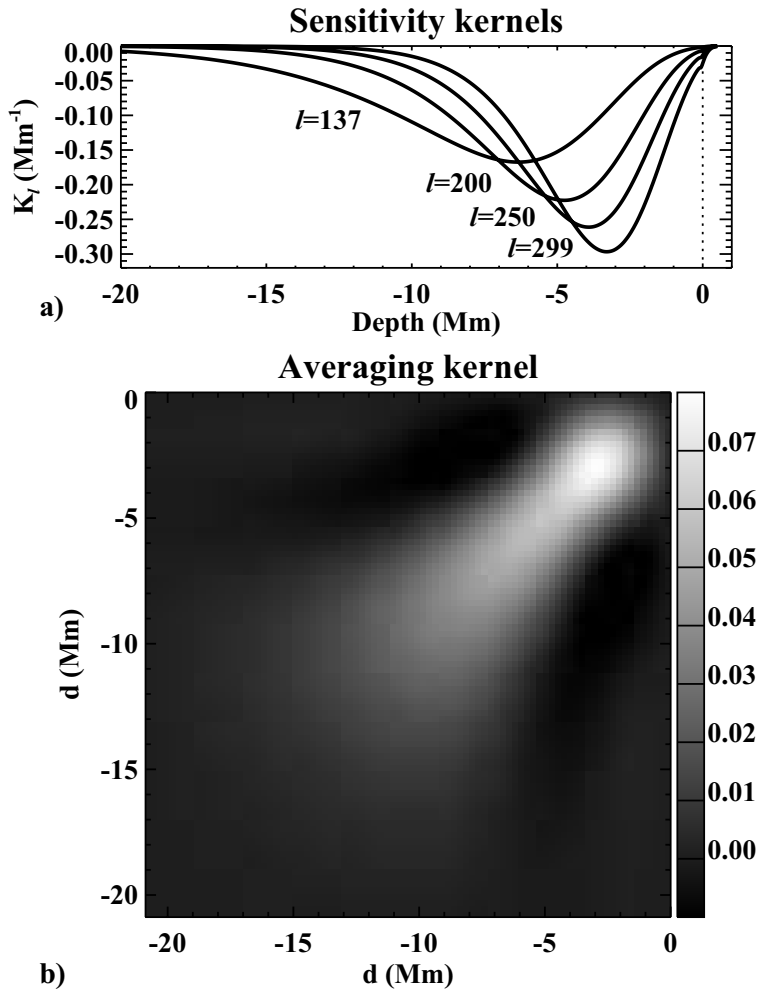}
	\caption{Variations with depth beneath the Sun's surface of: a) the sensitivity kernels, $K_\ell$, for a sample of angular degrees: $\ell$=137, 200, 250, and 299. b) The averaging kernel matrix showing the spatial resolution of the data inversion.}\label{fig3}
\end{figure}

However, the decrease of $\left<\Delta R/R\right>$ does not represent simple "shrinking" of the Sun. In fact, the different subsurface layers are displaced by different amount, and these variations are not necessarily homologous. To determine these variations we adopt the helioseismology inversion approach \citep{Lefebvre2005}. According to this method the depth dependence of the radial displacements, $\delta r/r$, is determined by solving the system of integral equations (Eq.~\ref{eq2}). Localization of the integral kernels, $K_\ell(r)=S_\ell(r)/I_\ell$, for different modes is different (Fig.~\ref{fig3}a), and this allows us to determine the depth dependence of $\delta r/r$. Eq.~\ref{eq2} is solved by applying the Tikhonov-Phillips \citep{Phillips1962,Tikhonov1963} regularization method in the formulation of \citet{Twomey1963}. 

Because the kernel functions are significantly different from zero only in the near-surface layers we solve Eq.~\ref{eq2} in the range of $[0.97R_\odot, 1R_\odot]$. In this interval we introduce an uniform radial grid, $r_i, i=0, ..., N-1$, and approximate $\delta r/r$ by a piece-wise linear function with unknown grid values $f_i=(\delta r/r)_i$. Then, Eq.~\ref{eq2} is reduced to a system of linear equations for $f_i$:
\begin{equation}
Af=g+\sigma,
\label{eq4}
\end{equation}
where elements of matrix $A$: $a_{i\ell}=\left[K^{(1)}_{il}-\left(K^{(2)}_{il}-K^{(2)}_{i-1,l}\right)\right]/\sigma_i$, $K^{(1)}_{il}=\int_{r_i}^{r_{i+1}}K_\ell(r)dr$, $K^{(2)}_{il}=\int_{r_i}^{r_{i+1}}K_\ell(r)(r-r_i)/(r_{i+1}-r_i)dr$, $g_\ell=\left(\frac{\Delta R}{R}\right)_\ell/\sigma_\ell$, $\sigma_\ell$ are measurement error estimates of the f-mode frequencies, which are used as a weighting function. 

Following  \citet{Twomey1963}, Eq.~\ref{eq4} is solved by minimizing the second derivative of $f$:
\begin{equation}
f=\left(A^*A+\gamma H\right)^{-1}A^*g,
\label{eq5}
\end{equation}  	
where $A^*$ is transposed of $A$, matrix $H$ represents the smoothness constraint, and $\gamma$ is the regularization parameter. It was chosen by applying the L-curve method of Hansen \citep{Hansen1992}.  For the expression of $H$ and computational details we refer to \citet{Twomey1963}. 

  The spatial resolution of the inversion results is estimated as half-width of averaging kernels given by matrix $\left(A^*A+\gamma H\right)^{-1}A^*A$ which is displayed in Fig.~\ref{fig3}b. It shows that the given set of f-modes provides localized solution in the depth range from 2 to 10~Mm. The uncertainties in variations of the seismic radius were estimated using a Monte-Carlo simulations: the inversion procedure was repeated 100 times by perturbing the f-mode frequencies with random Gaussian noise with the standard deviation corresponding to the observed error estimates of the SDO JSOC data archive.

\begin{figure}
	\centering
	\includegraphics[width=10cm]{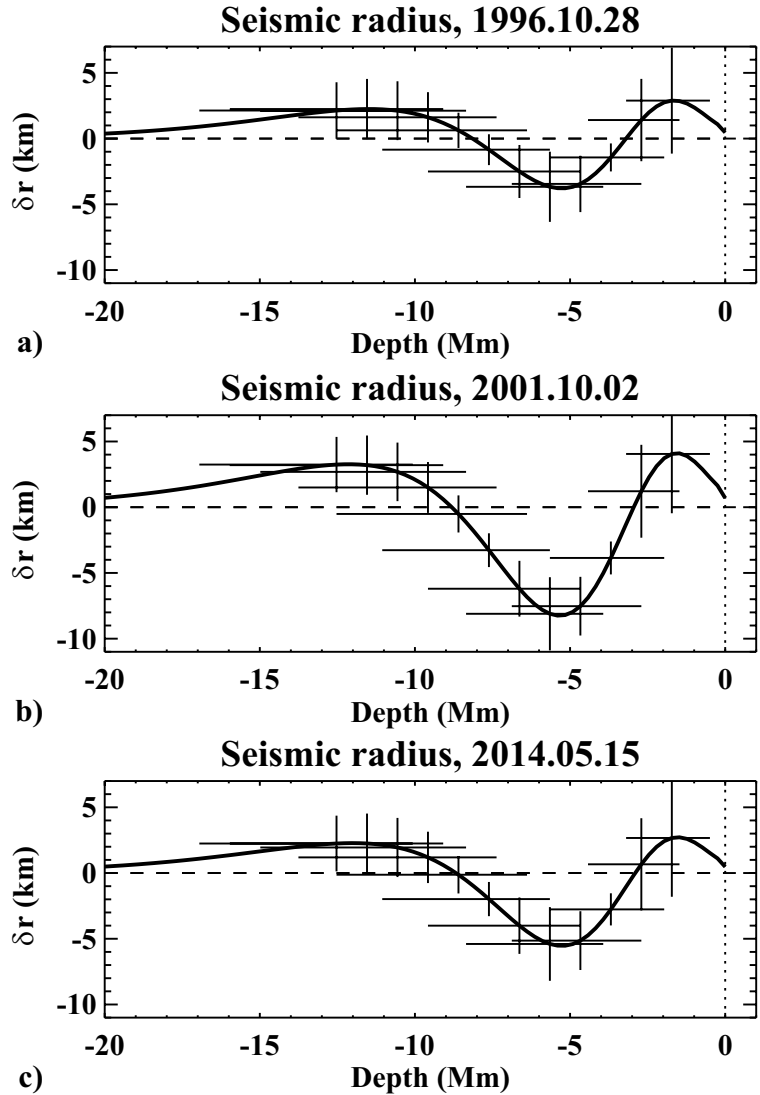}
	\caption{Variations with depth of the seismic radius relative to the reference period of the solar minimum in 2009 for: a) the beginning of Solar Cycle 23, b) maximum of Cycle 23, and c) maximum of Cycle 24. The error bars are $\pm 1\sigma$ uncertainties estimated by performing Monte-Carlo simulations. The horizontal bars show half-width of the averaging kernels.}\label{fig4}
\end{figure}

\begin{figure}
	\centering
	\includegraphics[width=10cm]{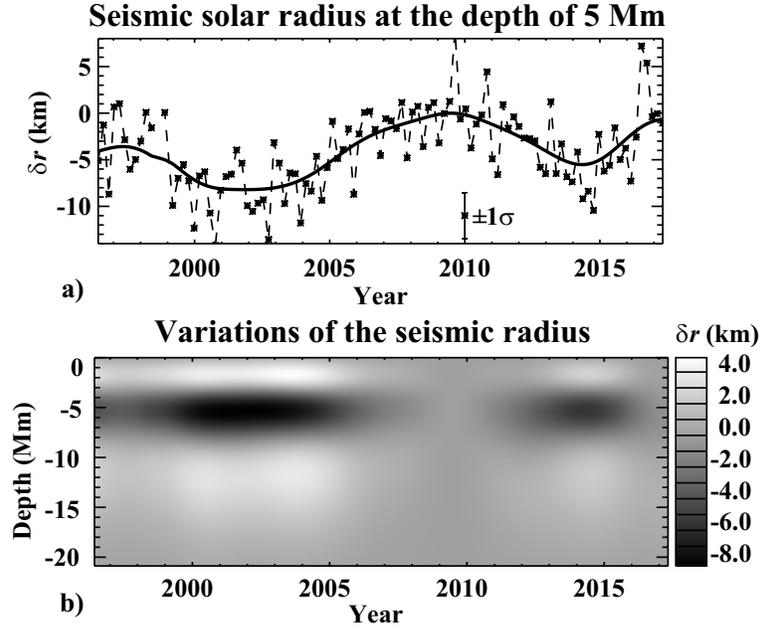}
	\caption{Variations of the seismic radius:  a) changes of the seismic radius at the depth of 5 Mm; b) variations with time and depth beneath the solar surface. The dark color (negative values) correspond to contraction and light color (positive values) to expansion of subsurface layers.}\label{fig5}
\end{figure}

 The inversion results for $\delta r$ shown in Fig.~\ref{fig4} for the three intervals corresponding to the beginning of Cycle 23 and the solar maxima of Cycles 23 and 24 reveal that most of the radius displacement was beneath the visible surface in the depth range of 3-8~Mm. The radius variations were not monotonic: above and below of this range the radius increased by 1-2~km. This means that the deeper layers (6-10~Mm) were compressed while the subsurface layers (3-5~Mm) expanded. The {  most significant} variations were at the depth of about {  $5\pm 2$}~Mm, and reached about {  $8\pm 3$}~km in Cycle 23 and {  $5\pm 3$}~km in Cycle 24 (Fig.~\ref{fig5}a). Figure~\ref{fig5}b shows the variations of displacement $\delta r$ (smoothed in time using a Gaussian kernel with the standard deviation of one year to remove the annual instrumental variations) as a function of depth during the whole 21-year period of the helioseismology observations from SoHO and SDO, relative to the deep activity minimum between Cycles 23 and 24 in 2010. Apparently, in Solar Cycle 23 the radius changed  substantially greater than in Solar Cycle 24.

\section{Discussion}

The presented analysis resolves the previous uncertainties in determination of variation of the seismic radius, caused by annual variations of observational data and potential instrumental effects. It takes into account variations of the f-mode frequencies, associated with the surface activity. The results show that the mean seismic radius estimated from the f-mode frequencies in the angular degree range of 130-300, decreased during the solar maxima by $\sim 3$~km in Cycle 23 and $\sim 2$~km in Cycle 24. The frequency inversion showed that the most significant changes occurred in the depth range of 3-8~Mm beneath the photosphere.   

Because f-mode frequencies are not sensitive to temperature or sound-speed variations the physical mechanism of the inferred displacements is probably associated with magnetic fields accumulating in the subsurface layers during the solar maxima. Currently, there is no theory to relate the inferred displacements to changes in properties of the solar magnetoconvection. Nevertheless, simple relations obtained by Goldreich et al. \citep{Goldreich1991} provide an interesting insight into properties of the subsurface fields \citep{Dziembowski2001}. According to this theory the Lagrangian change of the local radius can be expressed in terms of the averaged temperature and magnetic field changes by using the hydrostatic equilibrium condition and thermodynamic relations:
\begin{equation}
\Delta r \sim L\left[\frac{\Delta(\beta P_m)}{P_g}+\frac{\Delta T}{T}\right], 
\label{eq6}
\end{equation}
where $\Delta r$ is the radius change over distance $L$, $P_m=\left(\overline{B_h^2}+\overline{B_r^2}\right)/8\pi$ is the magnetic pressure, $\beta=\left(\overline{B_h^2}-\overline{B_r^2}\right)/(8\pi P_m)$
is a measure of the anisotropy of the field, $P_g$ is the gas pressure and $\Delta T/T$ is the relative temperature change. For a pure radial magnetic field $\beta=-1$, and for an isotropic magnetic field $\beta=1/3$. Therefore, the radius may decrease due to a local decrease of temperature or due a predominantly radial magnetic field.  To explain the radius decrease by 5-8~km in a 10~Mm subsurface layer the relative temperature change should be $\sim -5\times 10^{-4}$. This could result in an increase in the superadiabatic gradient, and a corresponding change in the energy flux and the solar irradiance. Alternatively, the radius change can be explained if the magnetic field is predominantly radial in the subsurface layer located at the depth 5-10~Mm beneath the solar surface. The gas pressure in this layer, $P_g \approx 2\times 10^9$~dyn~cm$^{-2}$. Hence, the average magnetic field strength is: $\sqrt{\overline{B_r^2}}\approx \sqrt{8\pi P_g\Delta r/L}\simeq 10$~kG. 

This should be considered as an upper limit on the field strength. As shown by numerical simulations \citep[e.g.][]{Chen2017,Kitiashvili2010} the subsurface magnetic field becomes predominantly vertical due to convective downdrafts that cause concentration of magnetic field into vertical structures of several kG strength. This is significantly greater than the strength of magnetic field emerging on the solar surface.

 More studies based on realistic numerical simulations of solar magnetoconvection and oscillations are needed for better understanding the observed variations of the f-mode frequencies, and for more accurate interpretation of  variations of the Sun's seismic radius with the activity cycle.

The work was performed with the support of the International Space Science Institute (ISSI) in Bern (CH), the VarSITI (Variability of the Sun and Its Terrestrial Impact) Program of the Scientific Committee On Solar-Terrestrial Physics (SCOSTEP). The authors thank the ISSI for holding a scientific meeting on solar variability organized by K. Georgieva. The work was partially supported by NASA grants  NNX14AB70G and NNX17AE76A.

\end{document}